# A hybrid solution for simultaneous transfer of ultrastable optical frequency, RF frequency and UTC time-tags over optical fiber

Przemysław Krehlik, Harald Schnatz, and Łukasz Śliwczyński

*Abstract—* We describe a fiber-optic solution for simultaneous distribution of all signals generated at today's most advanced time and frequency laboratories, i.e. an ultrastable optical reference frequency derived from an optical atomic clock, a radio frequency precisely linked to a realization of the SI-Second, and a realization of an atomic timescale, being the local representation of the virtual, global UTC timescale. In our solution both the phase of the optical carrier and the delay of electrical signals (10 MHz frequency reference and one pulse-per-second time tags) are stabilized against environmental perturbations influencing the fiber link instability and accuracy. We experimentally demonstrate optical transfer stabilities of $5\times10^{-18}$ and $2\times10^{-15}$ for 100 s averaging period, for optical carrier and 10 MHz signals, respectively.

## I. INTRODUCTION

FOR today's time and frequency (T&F) metrology the generation of precise and accurate frequency references or timescales are important for the most advanced scientific and industrial applications such as the search for a temporal variation of fundamental constants, geodesy, or mobile telecommunication.

The references or timescales discussed here are derived from either primary clocks such as Cs or Rb fountain clocks operating in the microwave domain [1], or more recently from optical atomic clocks, such as lattice clocks or single ion clocks [2]-[6]. Primary clocks developed over the last six decades have reached an uncertainty level in the low $10^{-16}$ [7], but optical atomic clocks supersede their microwave counterparts by two orders of magnitude reaching down to a few times $10^{-18}$. Considering this and the fact that optical clocks are becoming operational in many laboratories worldwide, the generation of a superior optical timescale has been discussed and has recently been demonstrated in [7].

However, the generation of such accurate frequency references and timescales requires substantial effort in terms of cost and personnel, thus is realized only in a few dedicated institutions over the world - usually National Metrology Institutes (NMIs). In this situation the distribution of locally generated reference signals to remote users such as universities, scientific or navigation centers, telecommunication institutions, potentially other (smaller) NMIs, and other demanding customers is crucial for their widespread use. The traditional solutions for long-haul T&F transfer are based on satellite techniques, that use either GNSS or geostationary satellites [8]-[10]. The satellite T&F transfer is widely accessible, but does not allow to exploit the actual accuracy and stability of the best atomic clocks and timescales at the remote end. Additionally, even sophisticated satellite techniques allow only to compare two distant clocks by long-term data acquisition and post-processing due to the inherent noise of the T&F satellite link. This hampers real-time applications that require ultra-low noise at short time scales.

Various techniques based on optical fibers as a transmission medium are a rapidly developing alternative for the distribution of time and/or frequency signals [11]-[14]. However, the main problem which has to be addressed in any solution are the phase (delay) fluctuations, induced in the fiber by environmental perturbations, such as temperature variations, vibrations, and mechanical stress. Additionally, for a transfer of a timescale (e.g. a UTC realization produced at a NMI), the signal propagation delay of the entire transmission system has to be stabilized and calibrated.

The common idea used in practically all fiber-optic T&F distribution systems is to arrange a bi-directional transmission in a single fiber, to take advantage of the high symmetry of the phase fluctuations occurring in both directions. The system may operate either as a simple two-way link or in a closed-loop phase-stabilizing scheme.

For transferring a radio-frequency (RF) reference signal (usually 10 MHz or 100 MHz) various solutions have been proposed [11], [15], [16]. Most of them are based on intensity-modulation (IM) and direct detection (DD) of the light propagating along the fiber link. In some cases they are combined with the transfer of a timescale (in form of a one-pulse-per-second - 1 PPS signal) [11], [17].

This work was supported by the EMPIR initiative co-funded by the European Union's Horizon 2020 research and innovation programme and the Participating States via project 15SIB05. Funding was received from the Faculty of Computer Science, Electronics and Telecommunications, AGH University of Science and Technology.

We would like to thank Thomas Legero for the help with providing ultrastable optical carrier used in experiments performed at PTB.

Przemysław Krehlik and Łukasz Śliwczyński are with Faculty of Computer Science, Electronics, and Telecommunications, AGH University of Science and Technology, Krakow, Poland (e-mail: krehlik@agh.edu.pl, sliwczyn@agh.edu.pl).

Harald Schnatz is with the Physikalisch-Technische Bundesanstalt, Braunschweig, Germany (e-mail: Harald.Schnatz@ptb.de).

For transferring a coherent optical frequency reference a continuous-wave (CW) optical carrier is disseminated via the fiber. It is derived from an ultrastable laser which is directly linked to the optical atomic clock by means of the optical frequency comb [18]. This guarantees a fixed, exactly known frequency ratio and phase relation between the optical clock and the CW laser.

Considering this, the question arises whether these two different approaches might be combined in a single transfer system to obtain a more versatile solution, offering distribution of both "traditional" electrical T&F signals and the most precise optical frequency references at the same time.

In [19] the authors proposed and demonstrated a system in which the optical carrier was phase-modulated by pseudorandom signals produced by a commercial two-way satellite time transfer modem, used routinely for comparing remote timescales [20]. In a proof-of-principle experiment [21], a periodically applied linear frequency chirp was applied to a transmitted, phase-stabilized optical carrier as a mean to determine the offset between timescales realized at both sides of a fiber link. In the experiment the timescales were represented by the 10 MHz signals referencing the frequency counters at both ends. In [22] and [23] the delay-stabilized transfer of multiple comb frequencies of an optical femtosecond frequency comb was proposed for simultaneously transferring signals in the optical and the RF domain.

In this paper we describe a novel solution of a hybrid system, that allows simultaneous dissemination of all the "T&F products" available in today's top T&F laboratories, i.e. a most stable optical frequency references derived from an optical atomic clock, an RF frequency reference precisely linked to a realization of the SI-Second, and a timescale realization UTC(k), being the local representation of the virtual, global UTC timescale. As a proof of concept (PoC) we present here an experimental realization of the system, together with an initial evaluation of the performance obtained. The conclusions and ideas for further development finalize the paper.

## II. CONCEPT OF THE HYBRID SYSTEM

The underlying idea of the solution proposed here is to combine the standard concept of the phase-stabilized optical carrier transfer [24], [25] with the ELSTAB technology [15] designed for delay-stabilized transfer of the RF frequency reference and pulse-per-second (1 PPS) time signal (see Fig. 1). The optical frequency stabilizing sub-system (marked on the figure with a red background color) is based on typical round-trip phase stabilization by means of the phase-locked loop (PLL). The part of the signal reaching the remote terminal is back-reflected towards the local terminal by a system composed of a Faraday mirror (FM) and a directional coupler C5, and is further combined at photodiode PD1 with the input optical reference. This produces a beat note that contains the information of the roundtrip phase fluctuations imposed by the fiber.

Using a digital PLL the phase of this beat note is compared with that of a voltage controlled oscillator (VCO) that drives an acousto-optic modulator (frequency shifter) AOM1 located at the input of the fiber link. In closed-loop arrangement the phase fluctuations imposed by the fiber are cancelled by the AOM1, provided that the phase fluctuations produced by the fiber are the same in both directions. Noting that AOM1 acts on forward and backward signals, it can be shown that also the optical carrier at the coupler C5 at the remote-side is stabilized. To separate the return optical signal from undesired reflections and backscattering a second acousto-optic modulator AOM2 is used at the remote end, which allows to apply a pass-band filter to the heterodyne beat note generated in the PD1 photodiode. (In our current design the both AOM frequencies are at +40 MHz, thus the PD1 beat note is 160 MHz.)

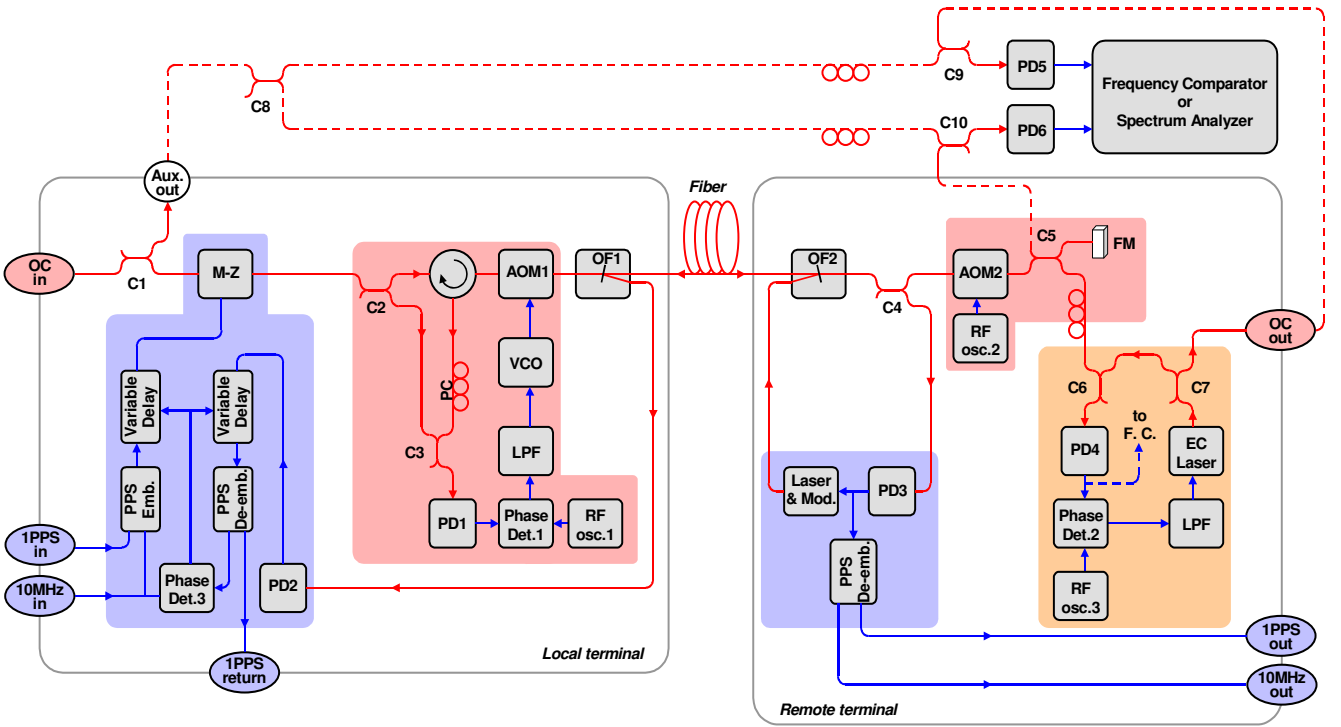

Fig. 1. Simplified block diagram of the proposed hybrid system. All PD modules comprise photodiode, electronic amplifier and, except PD2 and PD3, a PLL-based tracking filter.

In the PoC setup both local and remote ends are placed in the same laboratory. Thus, we added additional photodiode modules marked as PD5 and PD6 to measure the beat note between input and output optical carrier, allowing to assess the out-off-loop stability of the entire transfer system.

The optical carrier stabilization setup is realized in a "fiber optic domain", with no any free space optical components mounted on an optical breadboard. This approach made the setup portable and relatively cheap. We spliced the components with short fibers, and where possible, the lengths of fibers were matched to reduce the impact of ambient temperature fluctuations. The setup was mounted in standard fiber-optic plastic trays, fixed on an aluminum plate and placed in a cartoon box for thermal and acoustic shielding.

Having stabilized the optical carrier we now add the distribution system for 10 MHz signal and 1 PPS tags, both derived from an atomic clock producing a local UTC realization.

The electrical 10 MHz and 1 PPS signals are delay-stabilized by the same concept as in our ELSTAB system. The group delay fluctuations occurring in the fiber link (having exactly the same background as phase fluctuations of the optical carrier) are sensed and compensated by a delay-locked loop (DLL), in which the round-trip phase of the 10 MHz reference frequency is stabilized by phase detector controlling the two identical electronic delay lines, placed in both (forward and backward) signal paths. Once again, provided the same behavior of the fiber in both directions and identical characteristics of the delay lines, the remote side 10 MHz output is stabilized [15]. At the PPS Embedder block the sinusoidal 10 MHz signal is coherently converted to 1 MHz square wave with 1 PPS markers embedded in a form of specific phase modulation, which allows simultaneous transfer of the timing information. The opposite processing is performed at the PPS De-embedders, both at the remote terminal and in the backward path of the local one. The locally extracted 1 PPS signal (marked in Fig. 1 as 1 PPS return) is used for measuring the round-trip delay and then calibrating the output offset of the transmitted timescale [15].

In the standard setup of the ELSTAB system we supply the modulated square wave signal to a laser driver, and transmit it by means of the on-off intensity modulation of the telecommunication-grade distributed-feedback (DFB) lasers (local and remote ones), that are wavelength-stabilized by means of temperature stabilization. Because of the chromatic dispersion occurring in fibers, the potential long-term drifts of DFB lasers are converted into group delay instability, which is not symmetric in forward and backward directions, and thus not compensated and causing a systematic offset of output 1 PPS tags and phase fluctuations of the output 10 MHz signal. In practice in our standard ELSTAB setup the lasers' wavelength wander is one of main factors limiting the long-term stability of T&F output signals [14].

In the hybrid solution, described herein, we take benefit from modulating the highly-stable optical carrier instead of using a DFB laser in the local terminal. The amplitude-modulation (AM) with the depth of 15% (equivalent to 33% intensity modulation and 3 dB extinction ratio) is performed by the $LiNbO_3$ Mach-Zehnder modulator. In the current PoC setup, depicted in Fig. 1, the backward optical signal carrying the 10 MHz and 1 PPS is formed in the remote terminal by re-transmitting the incoming signal with the 100 GHz-shifted DFB laser, similarly as in the standard ELSTAB setup. Such relatively large frequency shift is necessary because the square wave modulated signal occupy wide bandwidth and thus should be separated from the other ones by means of optical filtering. Two three-port optical filters (add-drop multiplexers), marked in Fig. 1 as OF1 and OF2, are added for this purpose.

In current PoC experiment the wavelength-stabilization of the laser for the backward direction is still based on the temperature stabilization only. However, the long term goal is to stabilize its wavelength relatively to the highly stable optical carrier that is available at the remote terminal. This way we will in future get rid of the problem of the long-term wavelengths wandering, occurring in the standard ELSTAB setup.

While this approach of combining two established techniques is straightforward, it imposes a significant imperfection: because of the simultaneous dissemination of the RF signals the optical carrier is AM modulated. This results in the modulation side-lobes spreading over a broad frequency range that will corrupt the spectral purity of the optical carrier at the remote-end output. In order to eliminate these undesired AM side-lobes (odd multiples of 1 MHz located around the carrier), we added an optical carrier clean-up block (marked by orange background color) at the remote end of our hybrid system.

The clean-up comprises a narrow-linewidth (3 kHz) RIO ORION™ external-cavity semiconductor laser module with a frequency tuning input, a photodiode module with a narrow-band tracking filter (PD4) and a PLL circuitry which locks the laser to the incoming optical carrier with a 30 MHz offset. This way we not only eliminate the AM modulation, but also "compensate" the fiber attenuation, producing constant, high-power output signal regardless of the actual attenuation of the link.

III. EXPERIMENTAL RESULTS

The experiments were performed in three stages. In the first one, undertaken in Krakow at AGH University, we built the basic parts of the system and performed initial tests and optimization. At this stage the optical carrier clean-up block was not implemented and we used the free-running RIO laser as a substitute for an ultrastable optical carrier source. The coherence length of our laser (app. 70 km) was too short to analyze the transfer stability in even moderate lengths links, nevertheless for a 60 km test link we observed reasonably low Allan Deviation of about $2\times10^{-16}/\tau$ for 0.1 s $<\tau<$ 10 000 s (5 Hz bandwidth) with some small environmental perturbation for $\tau$ around a few minutes.

In the next stage, the setup was transported to the Physikalisch-Technische Bundesanstalt (PTB), Braunschweig, where the remote optical carrier clean-up block was implemented and the signal derived from an ultrastable laser was provided to the optical carrier input. The ultrastable laser comprises a fiber laser operating at 194 THz, a high finese optical cavity (F=316 000) and an electronic feedback system based on the standard Pound Drever Hall locking [26]. The ultrastable laser has a linewidth of app. 1 Hz and a residual drift of 17 kHz per day.

In the first step we measured the efficiency of the clean-up sub-system by comparing the power spectral density of the local-remote carrier beat note both for the output without clean-up (signals from C1 and C5 couplers provided for PD6 in Fig. 1), and with the clean-up (signals from C1 and C7 provided for PD5). One may notice (see Fig. 2) that the AM side lobes are eliminated by the clean-up subsystem as intended. Additionally, one may notice the significantly lower level of the broadband noise for the beat note with the clean-up. It is related to the fact that the power of the signal outgoing the clean-up laser is high enough for shot-noise-limited detection, which was not the case for the weak signal extracted from the fiber link.

In the next step we measured the stability of the remote optical carrier with a dead-time free multichannel frequency counter, both for the direct output (PD6 beat note) and for the clean-up sub-system (PD5 beat note). Additionally, we analyzed the in-loop beat note of the clean-up PLL, i.e. the output of PD4. The measurements were repeated for three cases: local and remote terminals connected with a short patchcord plus 10 dB attenuator, next with 50 km-long fiber (on a spool), and finally with 100 km-long fiber. (Please note that spooled fiber is much more exposed on acoustic noise than a field-deployed cable.)

We performed two sets of measurements: first, when the Mach-Zehnder modulator was driven by ELSTAB and the optical carrier was AM modulated, and second, without modulation. The modified Allan Deviation (MADEV) for a 100 km-long fiber and the AM modulation turned on is shown in Fig. 3.

The instability of the local-remote beat note with no clean-up (at PD6, green curve) starts slightly above $1\times10^{-17}$ at 1 s averaging, following a slope of $\tau^{-3/2}$ for averaging up to about 30 s. For long averaging the slope decreases which we attribute mainly to temperature fluctuations of the small residual uncompensated fibers.

The instability of the cleaned-up output (red curve) is about one order of magnitude higher. It starts slightly above $1\times10^{-16}$ at 1 s averaging, following a slope of $\tau^{-3/2}$ for averaging up to about 30 s. We attribute the significantly smaller slope for $\tau > 30$ s mainly to the temperature fluctuations of the significantly longer uncompensated fibers (app. 2 m) necessary for providing the signals to the coupler C9 producing the beat note at PD5. (This is an accidental feature of our actual setup, which had undergone several modifications.)

The main limiting factor (for short averaging, $\tau < 30$ s) for the achievable instability in the current implementation is due to the significantly higher frequency noise of the free-running clean-up laser (several kHz) as compared to the ultrastable master laser and its limited tracking bandwidth.

For comparison the in-loop beat stability of the clean-up sub-system (blue curve) is shown. Being identical with the out-of-loop signal for $\tau< 30$ s, it continuous with a slope of $\tau^{-3/2}$ for longer averaging as expected for a phase locked system.

The results shown in Fig. 3 were registered over the weekend, when there was no additional activity in the laboratory. The measurement repeated during the working days show a bit worse results for long averaging times; the MADEV at $\tau \approx 10^4$ s was $1\times10^{-19}$. This supports the conclusion that environmental noise acting on uncompensated fibers is the dominant limitation for long term measurements.

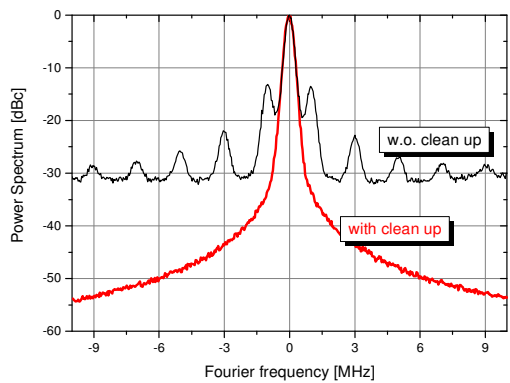

Fig. 2. Clean up efficiency. The analyzer bandwidth was 300 kHz. 100 km of fiber on spools was placed between local and remote terminals. Central frequencies of the beat notes (50 MHz with clean-up and 80 MHz without) are subtracted on x-axis.

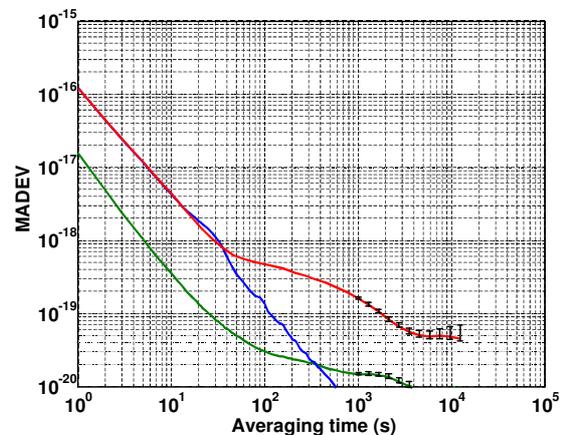

Fig. 3. ModADEV plots for beat notes at clean-up output (red), direct output (green) and in-loop signal in clean-up PLL (blue).

We did not observe any noticeable difference for the case when the modulation was turned off. From this we conclude that the AM carrier modulation in our hybrid system does not affect noticeably the fiber noise cancellation system.

The similar measurements performed for a shorter fiber span (50 km spool) and without the fiber show no significant difference in terms of stability for local-remote beat note when the clean-up sub-system was involved. For short averaging times the results were identical, for long averaging there were some minor differences related to probably unstable environmental conditions (temperature and human activity in the lab). A noticeable, but expected difference was observed however for the beat note measured without clean-up, where the short-term stability was better for shorter fiber ($9\times10^{-18}$ at 1 s), and the best for no fiber ($1.1\times10^{-18}$ at 1 s).

Overall, we conclude that the achieved performance does not impose a significant limitation even for modern optical clocks [27].

We identified however room for improvement in the short term stability, where the noise of the clean-up laser is certainly a limiting factor. Currently we use a frequency tuning input of the commercial RIO laser module, which has a limited bandwidth and this affects overall performance of the cleaning PLL. In future the bandwidth of this relatively slow tuning might be extended by implementing a "fast" tuning path, based on an acousto-optic modulator, or a custom electronic circuitry for laser current control. The medium and long term stability might be possibly improved by better thermal isolation and mechanical stiffness of the setup, and mainly by shortening the fibers involved in interferometer and providing signals for the beat notes. Particularly, the fibers providing the signals for PD5 are relatively long, as they were not included in the initial setup and were added later.

During the limited period of the experiments performed at PTB we focused on optical carrier measurements. Thus 10 MHz and 1 PPS signals stability was only monitored from time to time, and their long-term analysis was continued in the next period, again at AGH University.

Back in Krakow we again used the free-running RIO laser as an optical carrier source, and connect the local and remote terminals with the 60 km-long outdoor fiber. We checked the stability of the optical transfer by measuring the stability of the PD6 beat note, using a Symmetricom 5120A, just to ensure the correct operation of the optical carrier stabilization. Next we measured the long-term stability of the 10 MHz system with the Symmetricom 5120A, a high-speed digital oscilloscope and a custom programmable PPS generator for time transfer stability analysis, similarly as in our previous works [14], [15], [28]. The results, shown in Fig. 4 and Fig. 5 are very closed to those obtained previously for the ELSTAB alone [28]. For the 10 MHz signal the MADEV was $3.5\times10^{-13}$ at 1 s and $2.5\times10^{-16}$ at $10^3$ s averaging, and the time deviation (TDEV) for the 1 PPS signal was well below 1 ps for the entire observation range.

Summarizing this part of experiment we conclude that the changes introduced to ELSTAB by hybridization are not essential for 10 MHz and 1 PPS stability. It is especially related to the substantially modified scheme of the forward optical signal. In the standard ELSTAB setup the local-terminal output signal is generated by a DFB laser modulated by an integrated electro-absorption modulator, while in the hybrid setup we applied the external Mach-Zehnder modulator. As a known feature of the Mach-Zehnder modulator is some temporal drift of its transfer characteristic related to the properties of $LiNbO_3$ material [29], we used a custom driver circuitry to compensate this drift. This driver circuit ensures a very constant propagation delay and transfer function of the modulator [30]. In this experiment we verified for the first time the performance of this driver in a fully functional setup.

## IV. Conclusion

In this paper we presented a combined system capable of transferring all "products" of the most advanced time and frequency laboratories: a 1 PPS signal traceable to UTC(k), a RF reference frequency and an optical reference frequency. The presented solution is a complete "three inputs - three outputs" system dedicated for on-line distribution of the input signals to a remote laboratory, where no additional instrumentation is required nor clocks have to be operated. Simultaneous delivery of highly stable and traceable optical and RF references allows the users to synthesize the signals in other bands of optical spectrum as well as in the microwave domain. Potential users might be laboratories involved in clocks/oscillators development, high-resolution spectroscopy, radioastronomy observations, relativistic geodesy and so on.

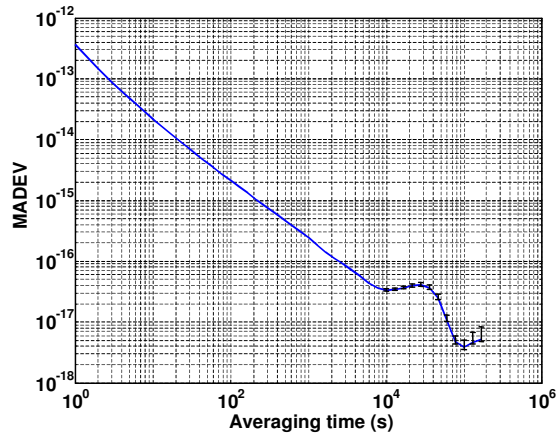

Fig. 4. 10 MHz stability for 60 km outdoor fiber.

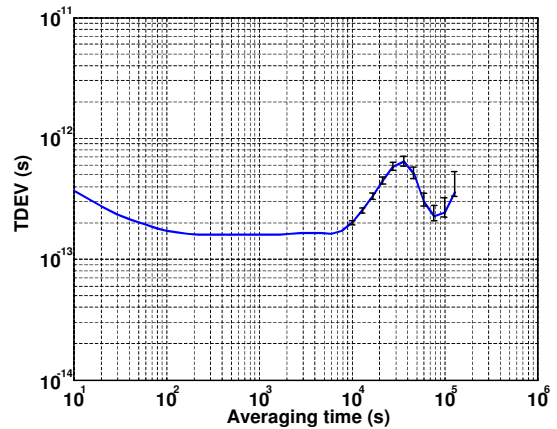

Fig. 5. 1 PPS stability for 60 km outdoor fiber.

The important feature of the proposed solution is that it is cost effective and easily transportable, as it comprises neither free space optics nor vacuum chambers for high finesse cavities, commonly used in optical frequency generation and dissemination systems.

The experimentally assessed stability of demonstrated dissemination system is $1.2 \times 10^{-16}$ at 1 s and $1.5 \times 10^{-19}$ at $10^3$ s for the optical carrier output, and $3.5 \times 10^{-13}$ at 1 s, $2.5 \times 10^{-16}$ at $10^3$ s for 10 MHz RF signal. The 1 PPS transfer stability is well below 1 ps for the entire observation range, i.e. up to $10^5$ s. Comparing this numbers with the typical stability of nowadays reference signals, we showed that this quality of dissemination does not impose noticeably degradation of the signals at the remote end.

The first PoC experiments described herein paves the way for further investigations and improvements. First, we noticed that the current clean-up subsystem based on commercial "black box" laser module with the vendor-implemented frequency tuning input limits the short-term stability of the cleaned-up output. Although we believe the current solution is quite satisfactory for majority of potential applications and really cost-effective, we are going to develop a locking system with higher bandwidth for the very same laser by RIO.

The next potential improvement is the mechanical (spatial) design of the fiber-optic setup used for stabilizing of output optical carrier; the fiber path starting at the input of AOM1 and ending at the Faraday mirror is stabilized, but thermal and acoustic perturbations affecting all other fibers involved in the optical carrier stabilization, cleaning-up and monitoring decrease the obtained results. In the current setup the "core" interferometer was built earlier, while the Mach-Zehnder modulator, the clean-up subsystem and the PD5 beat note arms were added later, which caused non-optimal spatial design and thus involved relatively long uncompensated fibers.

In future we are also planning to refer the wavelength of the backward laser in the ELSTAB remote module to the optical carrier. This way we will fully benefit from local-remote high stability of the optical carrier and reduce substantially the impact of fiber chromatic dispersion on RF signal stability and uncertainty of the time transfer.

Finally, we plan the experiment with much longer link (few hundreds kilometer) with bidirectional Erbium-doped optical amplifiers, which were earlier demonstrated as useful both in optical carrier dissemination [12] and ELSTAB installations [14].